\documentclass[preprint]{elsarticle}

\usepackage[english]{babel} 

\usepackage{amsmath,amsfonts,amsthm} 
\usepackage[mathcal]{eucal}

\usepackage[svgnames]{xcolor} 

\usepackage[hang, small, labelfont=bf, up, textfont=it]{caption} 

\usepackage{booktabs} 
\usepackage{lastpage} 

\usepackage{graphicx} 

\usepackage{enumitem} 
\usepackage{sectsty} 

\usepackage{graphicx}

\newtheorem{theorem}{Theorem}[section]

\usepackage{color,soul}
\usepackage{algorithmicx}
\usepackage{mathtools}
\usepackage{algpseudocode}
\usepackage[titletoc]{appendix}
\usepackage{color,soul}
\usepackage{enumitem}
\usepackage{hyperref}

\journal{Neurocomputing}

\begin{document}

\begin{frontmatter}

\title{Computing optimal discrete readout weights in reservoir computing is NP-hard}

\author[mymainaddress]{Fatemeh Hadaeghi \corref{mycorrespondingauthor}}
\cortext[mycorrespondingauthor]{Corresponding author}
\ead{f.hadaeghi@jacobs-university.de}

\author[mymainaddress]{Herbert Jaeger}

\address[mymainaddress]{Department of Computer Science and Electrical Engineering, Jacobs University Bremen, 28759 Bremen, Germany}

\begin{abstract}
We show NP-hardness of a generalized quadratic programming
problem, which we called \emph{unconstrained n-ary quadratic
programming} (UNQP). This problem has recently become practically relevant in the context of novel memristor-based neuromorphic microchip designs,
where solving the UNQP is a key operation for on-chip training of the neural network implemented on the chip. UNQP is the problem of finding a vector $\mathbf{v} \in S^N$ which
minimizes $\mathbf{v}^T\,Q\,\mathbf{v} +\mathbf{v}^T \mathbf{c} $,
where $S = \{s_1, \ldots, s_n\} \subset \mathbb{Z}$ is a given set
of eligible parameters for $\mathbf{v}$, 
$Q \in \mathbb{Z}^{N \times N}$ is positive semi-definite, and
$\mathbf{c} \in \mathbb{Z}^{N}$. In memristor-based neuromorphic hardware,
 $S$ is physically given by a finite (and small) number of possible memristor states. 
The proof of NP-hardness is by reduction from the \emph{unconstrained binary quadratic programming problem}, which is
a special case of UNQP where $S = \{0, 1\}$ and which is known to be
NP-hard. 

\end{abstract}


\end{frontmatter}


\section{Introduction}
The accustomed, apparently unbounded growth rate of digital computing
technologies begins to show signs of flattening out toward a ceiling
(``end of Moore's law''). This is due to several reasons, among them
the approaching of ultimate thermodynamical limits, the steeply
growing costs of building foundries for the respective next-generation
CMOS microchips, and technological difficulties in mastering device
miniaturization. Furthermore, the energy hunger of classical digital
computing technologies is increasingly becoming problematic, both for
global energy management and for deploying ever more computing-intense
AI algorithms on battery-powered personal devices. All of this has
revived interest in ``unconventional'' computing research, with
regards to non-digital material substrates, architectures and
algorithms \cite{Broersmaetal17}. Among the wide diversity of
approaches to unconventional computing, a leading role is emerging for
non-digital implementations of artificial neural networks (ANNs). Two
main trends in this arena are the exploit of low-energy, spiking
neural dynamics for ``deep learning'' solutions \cite{Tsaietal17}, and
``reservoir computing'' (RC) methods \cite{JaegerHaas04}. In material
RC implementations, a physical medium called the \emph{reservoir} ---
it can be electronic [13], optical \cite{Freibergeretal17},
nano-mechanical \cite{Coulombeetal17}, macro-mechanical
\cite{Nakajimaetal15}, or other [20] --- is nonlinearly excited by
temporal input signals, and from the resulting, high-dimensional
response signals within the medium a target output signal is delivered
through a trainable readout mechanism. The work reported in this
article arose within this latter line of investigation. Specifically,
we are involved in a European collaborative project (NeuRAM3,
\url{www.neuram3.eu}) which is concerned with the design of
memristor-based, spiking neuromorphic microchips. The reservoir here
is an analog VLSI recurrent neural network (RNN) --- concretely, the
currently available DYNAPs \cite{Moradietal18} and its future
descendants. The training of the readout mechanism amounts to solving
a linear regression problem, where the target output is a trainable
linear combination of the neural signals within the reservoir. Linear
regression solutions find the combination weights which minimize the
mean squared difference between the combined output signal and the
reference target. Linear regression is easily solved by standard
linear algebra algorithms when arbitrary real-valued combination
weights are admitted. However, for on-chip learning, the weights will
be physically realized by states of memristive ``synapses'', which
currently can be reliably set only to a very small number of discrete
values. This situation has led us to investigate the computational
nature of a ``discrete'' linear regression.

This article details our finding that here we are facing an
NP-complete problem. In Section 2 we formalize this problem as an
\emph{unconstrained n-ary quadratic programming} (UNQP) problem and
describe a known NP-complete problem, \emph{unconstrained binary
	quadratic programming problem} (UBQP). In Section 3 we show that
UBQP can be reduced to UNQP in polynomial time, thereby demonstrating
that UNQP is NP-hard.

\section{Problem statement}

In what follows, vectors are  column vectors unless otherwise defined. 

Reservoir computing is an approach to train recurrent neural networks (RNNs) in temporal signal processing tasks. The objective is to find a model RNN which approximates the input-output relation presented through matched training input-output signals $\mathbf{u}_{\mbox{\scriptsize \sf train}}(i), {y}_{\mbox{\scriptsize \sf train}}(i)$, $i = 1, \ldots, L$. 

The update equations of an elementary, discrete-time RC neural network are
\begin{equation}\label{eq1}
\mathbf{x}(i+1) = f(\mathbf{W}\mathbf{x}(i)+\mathbf{W}^{in}\mathbf{u}(i+1))
\end{equation}
\begin{equation}\label{eq2}
{y}(i) = (\mathbf{w}^{out})^T \, \mathbf{x}(i),
\end{equation}
where $i$ is discrete time, $\mathbf{u}(i) \in \mathbb{R}^{K}$ is the input signal, $\mathbf{x}(i) \in \mathbb{R}^{N}$ is the reservoir state, $f$ is a sigmoid function applied element-wise to its argument vector,  $\mathbf{W}$, $\mathbf{W}^{in}$ are the  recurrent and input weight matrices of size $N \times N, N \times K$, ${y}(i) \in \mathbb{R}$ is the output signal which is obtained by linearly combining the components of the reservoir state $\mathbf{x}(i)$ with the output weights $\mathbf{w}^{out}$ (an $N$-dimensional vector). Numerous variants and extensions of this basic system are being considered in RC, but for our current purpose this simple model with a scalar output signal suffices. 

Training an RC network proceeds in three stages:
\begin{enumerate}
	\item Create random recurrent and input weight matrices $\mathbf{W}$ and $\mathbf{W}^{in}$. 
	\item Drive the network \eqref{eq1} with the training input $\mathbf{u}_{\mbox{\scriptsize \sf train}}(i)$, obtaining reservoir response signals $\mathbf{x}_{\mbox{\scriptsize \sf train}} (i)$.
	\item Compute output weights $\mathbf{w}^{out}$ which minimize a loss function $$\mathcal{L}(\{(\mathbf{w}^{out})^T \, \mathbf{x}_{\mbox{\scriptsize \sf \sf train}} (i), {y}_{\mbox{\scriptsize \sf train}}(i)\}_{i = 1,\ldots,L})$$.
\end{enumerate}

The most popular loss function by far used in RC is the quadratic loss, which leads to solutions $\mathbf{w}^{out}$ that solve the linear regression problem
\begin{equation}\label{eq3}
\begin{aligned}
& \underset{\mathbf{w}}{\text{minimize}}
& & \| \mathbf{w}^T\, \mathbf{X} - \mathbf{y} \|^2,\\
\end{aligned}
\end{equation}
where $\mathbf{X} = (\mathbf{x}_{\mbox{\scriptsize \sf \sf train}} (1), \ldots, \mathbf{x}_{\mbox{\scriptsize \sf \sf train}} (L))$ and $\mathbf{y} = ({y}_{\mbox{\scriptsize \sf \sf train}} (1), \ldots, {y}_{\mbox{\scriptsize \sf \sf train}} (L))$.

In the RC learning paradigm,  $\mathbf{w}^{out}$ is the only item which is trained. The randomly created parameters in $\mathbf{W}$ and $\mathbf{W}^{in}$ are not adapted.  This latter condition has rendered RC interesting for unconventional hardware realizations of RNN-like learning systems, because in principle it allows one to employ any kind of nonlinearly excitable physical medium to instantiate the ``reservoir'' \eqref{eq1}.

In almost all currently realized physical reservoir computers, the output weights $\mathbf{w}^{out}$ are represented in a classical digital computer outside the unconventional physical reservoir. This means that floating-point precision numbers can be used in $\mathbf{w}^{out}$, and standard linear regression algorithms can be called upon. However, a widely considered goal for further development of unconventional RC systems is to integrate the output weights into the unconventional physical substrate. Much of today's unconventional-substrate RC research is concerned with neuromorphic hardware where the ``synaptic'' parameters contained in $\mathbf{W}, \mathbf{W}^{in}, \mathbf{w}^{out}$ are realized by memristors. It is currently infeasible to fabricate or tune memristive weights to anything approaching floating-point precision. In fact, one  has to face extreme low-precision scenarios, where realizable values of the elements in these matrices admit only ternary settings (for instance, taking approximate values $\{-s, 0, +s\}$) or other $n$-ary ranges $S = \{s_1, \ldots, s_n\}$ with for some quite small $n$. 

It is easily derived that the optimization problem \eqref{eq3} is equivalent to
\begin{equation}\label{eq4}
\begin{aligned}
& \underset{\mathbf{w}}{\text{minimize}}
& & \mathbf{w}^T Q \mathbf{w} +   \mathbf{w}^T \mathbf{c},\\
\end{aligned}
\end{equation} 
where $Q = \mathbf{X} \mathbf{X}^T$ and $\mathbf{c} = - 2 \, \mathbf{X} \mathbf{y}^T$. This is the format in which the linear regression objective is written in optimization theory contexts, where it is called the \emph{quadratic programming problem}.

In the unconventional RC hardware scenarios that motivated our investigation, the parameters admissible to be used in the components of $\mathbf{w}$ are constrained to a finite set $S = \{s_1, \ldots, s_n\}$ of $n$ real numbers. Our aim is to show that the resulting constrained quadratic programming problem is hard in the sense that it does not admit a general polynomial-time solution algorithm (provided that P $\neq$ NP). In order to show this, it suffices to show hardness for parameters in $Q, \mathbf{c}, S$ restricted to the integers $\mathbb{Z}$. That is, we consider the following optimization problem, which we will call the \emph{unconstrained n-ary quadratic programming (UNQP) problem}     
\begin{equation}\label{eq5}
\begin{aligned}
& \underset{\mathbf{w}}{\text{minimize}}
& & \mathbf{w}^T Q \mathbf{w} +   \mathbf{w}^T \mathbf{c},\\
& \text{subject to}
& & \mathbf{w} \in S^N,\\
\end{aligned}
\end{equation} 
where $Q \in \mathbb{Z}^{N \times N}$ is positive semi-definite, $
\mathbf{c} \in \mathbb{Z}^{N}$, $S = \{s_1, s_2, ..., s_n\} \subset
\mathbb{Z}$ and $|S| \geq 2$. 

Notice that UNQP is not a single optimization problem but a family
of such problems. For every choice of $S$, we obtain a distinct
optimization problem. We will show that for every such choice, the
resulting problem is NP-hard.   

\section{UNQP is NP-hard}

We now give a proof of
\begin{theorem} The problem UNQP from Equation \eqref{eq5} is NP-hard.
\end{theorem}

The proof is by transformation from the \emph{unconstrained binary quadratic programming (UBQP) problem} for vectors $\mathbf{v}$
\begin{equation}\label{eq6}
\begin{aligned}
& \underset{\mathbf{v}}{\text{minimize}}
& & \mathbf{v}^T Q \mathbf{v} +   \mathbf{v}^T \mathbf{c},\\
& \text{subject to}
& & \mathbf{v} \in \{0,1\}^N,\\
\end{aligned}
\end{equation} 
where  $Q \in \mathbb{Z}^{N \times N}$ is positive semi-definite and $\mathbf{c} \in \mathbb{Z}^{N}$. This problem  is known to be NP-hard \cite{pardalos1992complexity}. 

We assume that $S$ is ordered, that is $s_1  < \cdots < s_n$. Let $\mathbf{s}_j$ (where $j = 1, \ldots, n$) denote the $N$-dimensional vector whose compenents are all equal to $s_j$. Consider the problem

\begin{equation}\label{eq7}
\begin{aligned}
& \underset{\mathbf{t}}{\text{minimize}}
&&	(\dfrac{\mathbf{t}-\mathbf{s}_{1}}{s_{2}-s_{1}})^{T} Q (\dfrac{\mathbf{t}-\mathbf{s}_{1}}{s_{2}-s_{1}}) + (\dfrac{\mathbf{t}-\mathbf{s}_{1}}{s_{2}-s_{1}})^T \mathbf{c} \\
& \text{subject to}
&& \mathbf{t} \in \{s_{1},s_{2}\}^{N}.\\
\end{aligned}
\end{equation}

This is equivalent to UBQP, because a solution $ \mathbf{t}$ becomes a solution $\mathbf{v}$ of \eqref{eq6} by replacing components $s_1 \mapsto 0, s_2 \mapsto 1$ and vice versa. The problem  \eqref{eq7} can be re-written as
\begin{equation}\label{eq8}
\begin{aligned}
& \underset{\mathbf{t}}{\text{minimize}}
& & \mathbf{t}^T \tilde{Q}\mathbf{t} + \mathbf{t}^T \mathbf{\tilde{c}} + D\\
& \text{subject to}
& & \mathbf{t} \in \{s_{1},s_{2}\}^{N},\\
\end{aligned}
\end{equation}	
where  $\tilde{Q}=\dfrac{Q}{(s_{2}-s_{1})^{2}}$, $ \mathbf{\tilde{c}} = \dfrac{\mathbf{c}}{s_{2}-s_{1}} - \dfrac{2\,Q\mathbf{s}_1}{(s_{2}-s_{1})^{2}}$, 
and $D = \dfrac{\mathbf{s}_{1}^{T}Q\mathbf{s}_{1}}{(s_{2}-s_{1})^{2}} - \dfrac{\mathbf{s}_{1}^T \mathbf{c}}{s_{2}-s_{1}}$. Since $D$ is a fixed offset, this minimization problem is equivalent to 
\begin{equation}\label{eq9}
\begin{aligned}
& \underset{\mathbf{t}}{\text{minimize}}
& & \mathbf{t}^T \tilde{Q}\mathbf{t} + \mathbf{t}^T \mathbf{\tilde{c}}\\
& \text{subject to}
& & \mathbf{t} \in \{s_{1},s_{2}\}^{N}.\\
\end{aligned}
\end{equation}	

If $|S| = 2$, that is $S = \{s_{1},s_{2}\}$, the conversion  $s_1
\mapsto 0, s_2 \mapsto 1$ reveals that this problem is equivalent to
UBQP and hence NP-hard. In the remainder we assume that $|S| \geq 3$.  

In order to expand $\{s_{1},s_{2}\}$ to $S$, we add to the objective function in \eqref{eq9} a penalty term 
\begin{eqnarray*}
	M \sum_{i=1}^{N} (t_{i}-s_{1})(t_{i}-s_{2}) & = & M \, (\mathbf{t}-\mathbf{s}_{1})^{T}(\mathbf{t}-\mathbf{s}_{2}) \\
	& = & M \, (\mathbf{t}^{T} I \mathbf{t} - \mathbf{t}^T (\mathbf{s}_1+\mathbf{s}_{2}) + \mathbf{s}_1^{T}\mathbf{s}_{2})
\end{eqnarray*}
where $M \in \mathbb{N}$. This penalty term is zero when all components of $\mathbf{t}$ take  values in $ \{s_{1},s_{2}\}$ and positive when some components take values in $S \setminus  \{s_{1},s_{2}\}$. Adding this term to \eqref{eq9} and dropping the fixed offset $\mathbf{s}_1^{T}\mathbf{s}_{2}$ leads to 
\begin{equation}\label{eq10}
\begin{aligned}
& \underset{\mathbf{t}}{\text{minimize}}
& & \mathbf{t}^{T}(\tilde{Q} + MI)\mathbf{t}  + \mathbf{t}^T (\mathbf{\tilde{c}} - M (\mathbf{s}_1+\mathbf{s}_2)) \\
& \text{subject to}
& &   \mathbf{t} \in S^{N}.\\
\end{aligned}
\end{equation}

We now determine a lower bound for the penalty scaling factor $M$ which ensures that solutions $\mathbf{t}$ of \eqref{eq10} contain only components $s_1$ or $s_2$. Note that such solutions are also solutions of \eqref{eq9} and hence by the replacement  $s_1 \mapsto 0, s_2 \mapsto 1$ solutions of UBQP. 

Introducing $ H(\mathbf{t}) = \mathbf{t}^{T}\tilde{Q}\mathbf{t} + \mathbf{t}^T\tilde{\mathbf{ c}}$ and $ G(\mathbf{t}) = \sum_{i=1}^{N} (t_{i}-s_{1})(t_{i}-s_{2})$, the objective function in \eqref{eq10} can be written as
$$F(\mathbf{t}) = H(\mathbf{t})+ M G(\mathbf{t}) - M \mathbf{s}_1^{T}\mathbf{s}_{2}.$$

If we denote any $\mathbf{t}$ that contains some  element other than $s_{1}$ and $s_{2}$ by $\mathbf{t}^{(b+)}$ and any $\mathbf{t}$ that contains only $s_{1}$ and $s_{2}$ by $\mathbf{t}^{(b)}$,  the task is to find $M$ such that for any $\mathbf{t}^{(b+)}$,

\begin{equation*}
F(\mathbf{t}^{(b+)}) > max \{F(\mathbf{t}^{(b)}\}. 
\end{equation*}

Since $G(\mathbf{t}^{(b)})=0$, the task is thus find $M$ such that for any $\mathbf{t}^{(b+)}$,
\begin{equation*}
H(\mathbf{t}^{(b+)})  + M G(\mathbf{t}^{(b+)}) -  M \mathbf{s}_1^{T}\mathbf{s}_{2}  > max \{H(\mathbf{t}^{(b)} )-  M \mathbf{s}_1^{T}\mathbf{s}_{2}\}, 
\end{equation*}
or equivalently, we have to find $M$ such that 
\begin{equation}\label{eq11}
H(\mathbf{t}^{(b+)}) + M G(\mathbf{t}^{(b+)}) > max \{H(\mathbf{t}^{(b)})\}
\end{equation}
for any $\mathbf{t}^{(b+)}$.

An upper bound for $max \{H(\mathbf{t}^{(b)})\}$ is obtained by
noting that, since $\tilde{Q}$ is positive semi-definite, all
eigenvalues of $\tilde{Q}$ are non-negative and $0 \leq
(\mathbf{t}^{(b)})^{T}\tilde{Q}\mathbf{t}^{(b)} \leq
\lambda_{max}(\tilde{Q}) ||\mathbf{t}^{(b)}||^{2}$ where
$\lambda_{max}(\tilde{Q})$ is the largest eigenvalue of
$\tilde{Q}$. Let ${s}^{*}$ be the element of $\{s_{1}, s_{2}\}$ with
highest absolute value and let $\mathbf{s}^{*}$ be the vector of
${s}^{*}$'s, then for any $\mathbf{t}^{(b)}$

\begin{equation*}
{(\mathbf{t}^{(b)})}^{T}\tilde{Q}\mathbf{t}^{(b)} \leq \lambda_{max}(\tilde{Q}) ||\mathbf{t}^{(b)}||^{2} \leq \lambda_{max}(\tilde{Q}) ||\mathbf{s}^{*}||^{2}
\end{equation*}

and
\begin{equation*}
(\mathbf{t}^{(b)})^T	\mathbf{ \tilde{c}} \leq |(\mathbf{t}^{(b)})^T| \; |	\mathbf{ \tilde{c}} |\leq |(\mathbf{s}^{*})^T| \; |\mathbf{ \tilde{c}}|,
\end{equation*}

where the operation $|\mathbf{z}|$ sets each element in a vector $\mathbf{z}$ to its absolute value. Thus, 

\begin{equation}\label{eq12}	
K :=  
\lambda_{max}(\tilde{Q})||\mathbf{s}^{*}||^{2} +
|\mathbf{\tilde{c}}^{T}| |\mathbf{s}^{*}| \geq H(\mathbf{t}^{(b)}) 	
\end{equation}

is an upper bound of $max \{H(\mathbf{t}^{(b)})\}$. 
We have to find $M$ such that, for any $\mathbf{t}^{(b+)}$,

\begin{equation*}
H(\mathbf{t}^{(b+)})+ M G(\mathbf{t}^{(b+)}) > K, 
\end{equation*}
or equivalently, such that for any $\mathbf{t}^{(b+)}$,
\begin{equation}\label{eq13}
M> \frac{K-H(\mathbf{t}^{(b+)})}{G(\mathbf{t}^{(b+)})}.
\end{equation}

\eqref{eq13} is well-defined since $G(\mathbf{t}^{(b+)}) > 0$. We
proceed to find a lower positive bound $L_G$ for $G(\mathbf{t}^{(b+)})$
and a lower bound $L_H$ for $H(\mathbf{t}^{(b+)})$.

Observing that we assumed $S$ to be ordered, an obvious positive
lower bound $L_G$ is given by taking for $\mathbf{t}^{(b+)}$ a vector
made only of $s_{1}$'s and 
$s_{2}$'s except one element  set to  $s_{3}$, resulting in $L_G =
(s_{3}-s_{1})(s_{3}-s_{2})$. 

For $L_H$, let ${s}^{**}$ be the member of
$S = \{s_{1}, ..., s_{n}\}$ with highest absolute value and
$\mathbf{s}^{**}$ the vector of all ${s}^{**}$'s. A positive upper bound $K'$
for $| H(\mathbf{t}^{(b+)}) |$ is  given (re-using \eqref{eq12}) as
$ K' = \lambda_{max}(\tilde{Q})||\mathbf{s}^{**}||^{2} + |\mathbf{
	\tilde{c}}^{T}| |\mathbf{s}^{**}|$. Consequently, $L_H = -K'$ gives a
lower bound of $H(\mathbf{t}^{(b+)})$. 

In summary, 
\begin{equation}\label{eq14}
M = \frac{K - L_H}{L_G} 
\end{equation}
ensures that solutions $\mathbf{t}$ of the optimization problem
\eqref{eq10} contain only components from $\{s_1, s_2\}$. 

It is easy to see that the transformation from \eqref{eq6} to
\eqref{eq10} with $M$ determined by \eqref{eq14} can be
accomplished in polynomial time.

Some core ideas used in this proof were motivated by \cite{galli2014compact}
and \cite{billionnet2007using}.

\section{Discussion}

We only showed that UNQP is NP-hard, but did not show that it is in
NP. If true, this may turn out to be surprisingly difficult to
demonstrate. Related quadratic programming problems have been proven
to be in NP only long after NP-hardness was established, and those proofs
are rather involved \cite{vavasis1990quadratic, del2017mixed}.

The binary problem UBQP has numerous applications. There exists an
extensive literature on exact and approximate methods capable of
producing practical solutions for this problem in a large variety of
circumstances (survey in \cite{kochenberger2014unconstrained}). Some
of the existing methods for coming to practical terms with UBQP may
turn out to be adaptable to our more general problem UNQP, but this
remains to be worked out when the occasion arises. With regards to
reservoir computing, initial investigations indicate that very simple
approximations already work quite satisfactorily.

\section{Acknowledgments}

This work was supported by European H2020 collaborative project
NeuRAM3 [grant number 687299]. Fatemeh Hadaeghi would also like to
thank Adam N. Letchford, who provided insight and expertise that
greatly assisted this research.



\bibliography{Manuscriptref.bib}

\end{document}